\documentclass[12pt]{article}
\usepackage{t1enc}
\usepackage[utf8]{inputenc}
\usepackage[english]{babel}
\usepackage{graphicx}
\usepackage{amssymb}
\usepackage{amsmath}
\usepackage{amsthm}
\usepackage{braket}

\def\UN{\mathbf{1}}

\def\UN{\mathbf{1}}

\def\qed{\ \vrule height 5pt width 5pt depth 0pt}
\def\cros{\raise1.9pt\hbox{$\scriptscriptstyle
          >$}\!\raise1.5pt\hbox{$\scriptstyle\triangleleft\,$}}

\def\l{{\lambda}}

\def\w{\wedge}

\theoremstyle{definition}
\theoremstyle{definition}
\theoremstyle{definition}
\theoremstyle{definition}
\newcommand{\noi}{\vspace{0.1in} \noindent}

\title{\bf Commutativity, comeasurability, and contextuality in the Kochen-Specker arguments}
\author{\textit{Gábor Hofer-Szabó}\thanks{Research Center for the Humanities, Budapest, email: szabo.gabor@btk.mta.hu}}
\date{}

\begin{document}
\maketitle

\begin{abstract}
If noncontextuality is defined as the robustness of a system's response to a measurement against other simultaneous measurements, then the Kochen-Specker arguments do not provide an algebraic proof for quantum contextuality.  Namely, for the argument to be effective, (i) each operator must be uniquely associated with a measurement and (ii) commuting operators must represent simultaneous measurements. However, in all Kochen-Specker arguments discussed in the literature either (i) or (ii) is not met. Arguments meeting (i) contain at least one subset of mutually commuting operators which do not represent simultaneous measurements and hence fail to physically justify the functional composition principle. Arguments meeting (ii) associate some operators with more than one measurement and hence need to invoke an extra assumption different from noncontextuality.
\vspace{0.1in}

\end{abstract}

%\pagebreak
%\tableofcontents

\section{Introduction: the main argument in brief}\label{Sec:Int}

The aim of this paper is to challenge the view that Kochen-Specker (KS) arguments provide an algebraic proof for quantum contextuality if noncontextuality is interpreted as the robustness of a system's response to a measurement against other simultaneous measurements.

As a start, it is worth discerning KS \textit{arguments} from KS \textit{theorems}. KS theorems are simply mathematical theorems in form of a coloring problem, while KS arguments are physical arguments devised to prove that quantum mechanics (QM) is contextual. The KS theorems start from a family of self-adjoint operators arranged on a \textit{hypergraph}\footnote{A generalization of a graph where an edge can connect any number of vertices.} such that the subsets of mutually commuting operators define the \textit{hyperedges}\footnote{A non-empty subset of vertices.} of the hypergraph.\footnote{See e.g. (Abramsky and Brandenburger, 2011),  (Cabello et al. 2014), and (Acín et al., 2015).} Two examples for such a hypergraph are the GHZ graph (on the left) and the Peres-Mermin graph (on the right). 
\begin{figure}[h]
\centerline{\resizebox{9cm}{!}{\includegraphics{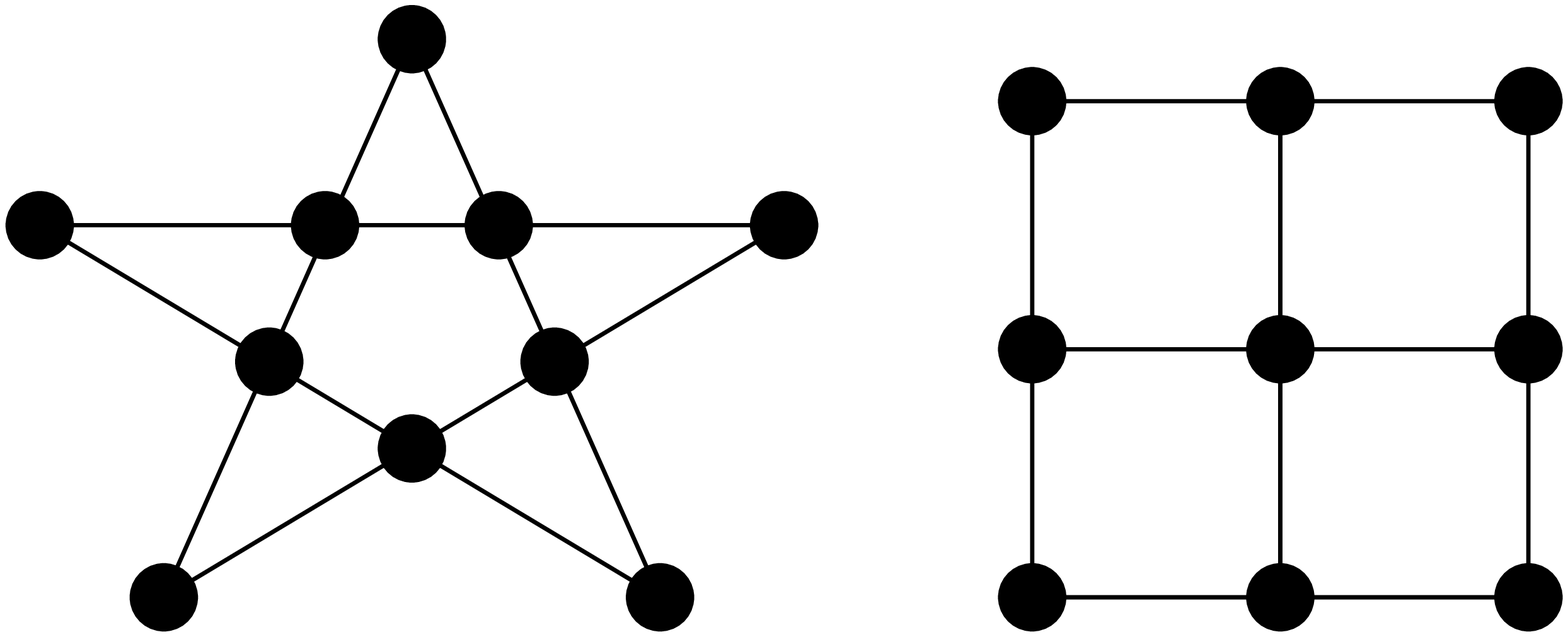}}}
\end{figure}
Here each hyperedge is depicted by an unbroken line connecting 4 collinear vertices on the GHZ graph and 3 collinear vertices on the Peres-Mermin graph. Next, one introduces \textit{value assignments} on the graph, that is, functions assigning to each vertex one of the eigenvalues of the operators represented by the vertex in every quantum state. Since the operators are typically projections or contractions, the assignments generally yield the numbers 0, $+1$ and $-1$. The value assignments are, however, constrained by the so-called \textit{functional composition principle}\footnote{See (Redhead, 1989, p. 121) and (Held, 2018, Sec. 4).} (FUNC) requiring that if the operators on a given hyperedge stand in a certain functional relation to one another, then the values assigned to the operators should also stand in the same functional relation in every quantum state.\footnote{Alternatively: the values assigned to mutually commuting operators are the eigenvalues corresponding to one of the common eigenstates of these operators.} In the case of the GHZ graph, for example, the product of the operators on every hyperedge is the unit operator $+\hat{\UN}$, except for the horizontal hyperedge, where the product is $-\hat{\UN}$. In the case of the Peres-Mermin graph the product of the operators on every hyperedge is $+\hat{\UN}$, except for the third vertical hyperedge, where it is $-\hat{\UN}$. Since the eigenvalues of each operator on both graphs is $\pm1$, FUNC allows for only such value assignments for which the product of the assigned numbers on every hyperedge equals the product of the operators (that is, $+1$ or $-1$) on that hyperedge. It is easy to show that there is no such value assignment on the above two graphs. More generally, KS theorems provide complex hypergraphs of operators such that there is no value assignment on the graph respecting FUNC. Some KS theorems work only in specific quantum states, others across all states. Thus, one can differentiate \textit{state-dependent} and \textit{state-independent (algebraic)} KS theorems.

To proceed from a KS theorem to a KS argument, one needs to provide a \textit{physical interpretation} for the KS graph. To this aim, one first assumes that QM admits an ontological (hidden variable) model. In other words, one assumes that the quantum states are simply distributions of underlying (dispersion-free) ontic states. Next, one associates the \textit{operators} with \textit{observables} and \textit{measurements}. Measurements are ``lists of instructions to be implemented in the laboratory'' (Spekkens, 2005, p. 2) and observables are physical magnitudes which characterize a given quantum system. In a value-definite (deterministic) ontological model each observable has a well-defined \textit{value} in every ontic state. Each observable is also associated with a measurement (procedure) such that the outcome of the measurement reveals (faithfully) the value of the observable. Furthermore, each observable $\mathcal{A}$ and the corresponding measurement $a$ is represented by a self-adjoint operator $\hat{\boldsymbol{a}}$ such that the values of the observable and the outcomes of the measurement are just the \textit{eigenvalues} of the operator. The exact nature of these associations will be examined below. Finally, one interprets the \textit{quantum probability} of an operator's spectral projection associated with a given eigenvalue as the \textit{probability} of the corresponding observables having the value associated with that eigenvalue, and also as the \textit{conditional probability} of the outcome associated with that value provided the corresponding measurement is performed.

On this interpretation each value assignment on a KS graph represents a possible distribution of values in a given ontic state which the observables associated with the operators on the graph can take and which the corresponding measurements reveal. The constraint FUNC is justified as follows. Mutually commuting operators on a hyperedge have common eigenstates. If one prepares the system in one of these eigenstates, then the functional relationship between the operators will be realized as the functional relationship between the outcomes of the corresponding measurements, and also between the values of the associated observables. Note that to justify FUNC in an eigenstate, the measurements need not be comeasurable (simultaneously measurable). But what justifies FUNC in a general quantum state? Here one can come up with three answers.

First, one can say that any ontic state featuring in the support of a general quantum state must also show up in the support of at least one eigenstate.\footnote{Maroney and Timpson (2014) call it ``operational eigenstate support macrorealism.''} This answer, however, is not very appealing. After all, why should every quantum state be composed of the same ontic states as the eigenstates are? 

Second, one can say that the mutually commuting operators $\{\hat{\boldsymbol{a}}_i\}$ of the graph represent \textit{simultaneous measurements} $\{a_i\}$ and on performing these joint measurements one can directly observe the functional relationship in question between the joint measurement outcomes and hence (assuming faithful measurement) between the values of the observables. Note that simultaneous measurements are understood here in the very physical sense, namely as measurements which can jointly be performed at the same time on the same system. Also note that, although simultaneous measurements get represented in QM by commuting operators, the converse is not true: from the mathematical fact that certain measurements are represented by commuting operators it does not follow that these measurements can be simultaneously performed. We come back to this important point below. 

Third, one can refer to the mathematical fact that for every set $\{\hat{\boldsymbol{a}}_i\}$ of mutually commuting operators sitting on a hyperedge there is an operator $\hat{\boldsymbol{b}}$ and functions $\{f_i\}$ such that $\hat{\boldsymbol{a}}_i = f_i(\hat{\boldsymbol{b}})$. Thus, one can say that there is only \textit{one single observable} $\mathcal{B}$ with a corresponding measurement $b$ and the set $\{\hat{\boldsymbol{a}}_i\}$ of mutually commuting operators simply represents the different functions $\{f_i(\mathcal{B})\}$ of this very observable. Consequently,  FUNC holds trivially: it simply expresses the functional relationship among the different functions of the outcomes of $b$. Note that in this case the measurements $\{f_i(b)\}$ associated with $\{\hat{\boldsymbol{a}}_i\}$ can be called ``simultaneously measurable'' only metaphorically since one performs only one single measurement, namely $b$, and applies the functions to the outcome. %Nevertheless, I will call these measurements simultaneous (sometimes putting the term in quotation marks).

Now we show that these latter two justifications of FUNC lead to two different realizations of a KS graph. To reduce metaphysics and to get closer to the experimental testability, we eliminate the concept of observable from the discussion and adopt an \textit{operational approach} relying purely on operators and measurements. We call an association of the operators of a KS graph with measurements a \textit{realization} of the graph. A realization is \textit{unique} if each operator on the graph is associated with only one measurement and \textit{non-unique} if some operators are associated with more than one measurement. A measurement associated with an operator is said to be \textit{realizing} the operator. Now, in the third justifications of FUNC above a set of operators $\{\hat{\boldsymbol{a}}_i\}$ sitting on a hyperedge is realized by one single measurement $b$ since the functions $f_i$ applied to the measurement $b$ are represented by $\hat{\boldsymbol{a}}_i$. Call a realization \textit{hyperedge-based} if there is at least one hyperedge on the graph which is realized by (different functions of) one single measurement. 

In a unique realization of the Peres-Mermin graph, for example, one has 9 different measurements associated with the 9 vertices (operators) of the graph. In a (maximally) hyperedge-based realization of the same graph one has only 6 measurements associated with the 6 hyperedges (three rows or and three columns) of the graph. Can this latter realization be unique? No, it cannot, as the following simple lemma shows:

\noi
\textit{Lemma.} A hyperedge-based realization in which all sets of mutually commuting operators represent simultaneous measurements cannot be unique. 

\noi
\textit{Proof.} Let $\hat{\boldsymbol{a}}_1$ be an operator sitting at the intersection of two hyperedges such that all operators (among them $\hat{\boldsymbol{a}}_1$) on the one hyperedge are realized by a measurement $b$. Suppose \textit{a contrario} that $\hat{\boldsymbol{a}}_1$ is realized only by $b$. Now, since mutually commuting operators represent simultaneous measurements, the measurements realizing the operators on the \textit{other} hyperedge must be comeasurable with at least one measurement realizing $\hat{\boldsymbol{a}}_1$. But there is only one measurement realizing $\hat{\boldsymbol{a}}_1$, namely $b$. Therefore, the measurements realizing the operators on the other hyperedge are comeasurable with $b$. But then all operators on the two hyperedges either represent functions of $b$ or measurements which are comeasurable with $b$. Assuming that simultaneous measurements get represented by commuting operators, this means that all operators on both hyperedges commute. Contradiction. Consequently, $\hat{\boldsymbol{a}}_1$ cannot be realized only by $b$. \qed

\noi
That is, a realization of a KS graph where all sets of mutually commuting operators are realized by simultaneous measurements but some such sets by one single measurement cannot be unique. In other words, only the above second justification of FUNC can lead to a unique realization, the third justification always leads to a non-unique realization.

To avoid the no-go result of the KS argument, unique and non-unique realizations follow \textit{different strategies}. On a unique realization one blocks the argument by assuming that at least one measurement (associated with an operator sitting at the intersection of two hyperedges) can have different outcomes in an ontic state depending on whether it is simultaneously performed with measurements represented by operators on the one or on the other hyperedge. On a non-unique realization, however, the argument can also be blocked by assuming that different measurements represented by the same operator (at the intersection of two hyperedges) can have different outcomes in a given ontic state.

These two strategies for avoiding the no-go result represent \textit{two different interpretations of (non)contextuality}. On the first interpretation, noncontextuality is the independence of the outcome of a measurement in every ontic state from which other measurements it is simultaneously measured with. On the second interpretation noncontextuality is a perfect correlation in every ontic state between the outcomes of two different measurements represented by the same operator.\footnote{Both definitions of noncontextuality can be generalized for \textit{probabilistic} ontological models by replacing ``outcome'' by ``probability distribution of the outcomes.''} Note that the two interpretations are different and logically independent. 

\noi
Historically, the first interpretation of noncontextuality goes back to Bell, the second interpretation to Van Fraassen. Bell  interprets noncontextuality as: the ``measurement of an observable must yield the same value independently of what other measurements may be made simultaneously'' (Bell, 1966/2004, p. 9). Van Fraassen's contextuality, however, is based on the insight that ``[t]wo observables [$a$ and $b$] are statistically equivalent if they have the same probability distribution \dots In that case they are represented in physics by the same Hermitean operator. \dots But that does not mean that $a=b$''  (Van Fraassen, 1979, p. 158). In other words, two observables can be represented by the same self-adjoint operator without being the same. But then, one is not forced to assign the same value to them. Redhead (1989, p. 135) calls this fact \textit{ontological contextuality}. 

Many authors working in the operational approach (Spekkens, 2005; Hermens, 2011; Leifer, 2014; etc.) follow this second interpretation. Spekkens, for example, writes: ``A noncontextual ontological model of an operational theory is one wherein if two experimental procedures are operationally equivalent [that is, they are represented by the same self-adjoint operator], then they have equivalent representations in the ontological model.'' (Spekkens, 2005, p. 1) There are also experiments devised to test noncontextuality in this second sense (Mazurek, 2016). The general idea behind this understanding of noncontextuality, once again, is that if two measurements---even if they are not simultaneous---are represented by the same self-adjoint operator (which, as Van Fraassen rightly says, empirically just means that the outcome statistics of the two measurement are the same), then it is rational to assume that in every ontic state the outcomes (or more generally, the probability distributions of the outcomes) of the two measurements are also the same.

I don't doubt that this is a reasonable requirement on an ontological model.\footnote{However, in Section \ref{Sec:Toy}, I show a simple classical ontological models in which this condition is violated.} I think, however, that this requirement is more closely related to the special way in which QM is representing the conditional probabilities and much less to the very concept of contextuality. If outcomes of different measurements (defined via different ``lists
of laboratory instructions'') are represented by the same projection, as happens in QM, then there might indeed seem to be a need for the ``context'' to dismantle what was put together by the representation. But this contextuality is simply the consequence of a special representation which does not discriminate mathematically between that which is different physically, namely the outcomes of different measurements. Had this difference been respected by the representation, ontological contextuality would not arise.

If one relies, however, on the everyday usage of the term, then ``context'' refers simply to the circumstances in which a certain event, observation or measurement  occurs. These circumstances are not constitutive in the definition of the very event or measurement, but can significantly influence the occurrence of the event or the result of the measurement. The important aspect of these circumstances, however, is that they are simultaneously present with the event or measurement. A possible context for a measurement in physics is another measurement which is performed simultaneously with the one in question. (A non-simultaneous measurement cannot provide such a context since it lives in another possible world.) In this sense noncontextuality refers to a kind of robustness of the definite response to a measurement on a given system, with respect to simultaneous measurements that are also performed on the system. I will refer to this kind of noncontextuality as \textit{simultaneous noncontextuality}. If we understand noncontextuality in this way, we just arrive at the above first interpretation of noncontextuality. 

I have no objection against using noncontextuality in the second sense as Spekkens and many others use it. However, in this paper I will use noncontextuality exclusively in the first sense (that is, as simultaneous noncontextuality) and refer to the second one as \textit{Spekkens' condition}. My aim is to explore \textit{whether the KS arguments can prove that QM is contextual in the first sense}. The challenge is then to construct (i) a unique realization for a KS graph, that is, to associate each operator of the graph with a different measurement such that (ii) mutually commuting operators represent simultaneous measurements. We stress that points (i) and (ii) are both important. Mutually commuting operators must represent simultaneous measurements, otherwise FUNC, on which the whole KS theorem is based, will not be physically justified. And the realization must be unique since non-unique realizations realizing certain operators by more than one measurement need to invoke noncontextuality in the second sense that is, Spekkens' condition. By abandoning Spekkens' condition (that is, by allowing the system to respond differently to different measurements represented by the same operator) one can always block the KS argument. In short, simultaneous measurability and unique realization are both \textit{sine qua non} in proving quantum contextuality.\footnote{Throughout the paper I will use the term ``quantum contextuality'' as the non-existence of a noncontextual value-definite ontological model for QM.}

\noi
In the paper I will proceed as follows. First, I introduce the framework of operational theories (Sect. \ref{Sec:Op}) and ontological (hidden variable) models (Sect. \ref{Sec:Ont}); and define (simultaneous) noncontextuality (Sect. \ref{Sec:NC}). Then, I accommodate QM in this framework (Sect. \ref{Sec:QM}); pick a simple example, the Peres-Mermin square (Sect. \ref{Sec:PM}); clarify what operational theories would realize it (Sect. \ref{Sec:OTPM}); and show that the standard spin measurement realization does \textit{not} do the job (Sect. \ref{Sec:SMPM}). Next, I categorize KS argument into three types (Sect. \ref{Sec:Three}), investigate the GHZ argument as an argument of type II (Sect. \ref{Sec:KS(ii)}); show that arguments of type III can be effective only if they switch to non-unique realization (Sect. \ref{Sec:KS(iii)}) and if they assume Spekkens' condition (Sect. \ref{Sec:Spek}). Using a simple toy model, I compare Spekkens' condition and noncontextuality (Sect. \ref{Sec:Toy}). Finally, I contrast the KS arguments with the Bell-type arguments (Sect. \ref{Sec:Conc}).

\section{Operational theories}\label{Sec:Op}

An operational theory is a physical theory specifying the probability of the outcomes of some measurements performed on a physical system prepared previously in certain states. Let $s, t,  ...\in S$ be the possible \textit{states} or \textit{preparations} of the system under investigation. Let $a, b, ... \in M^b$ be the \textit{basic measurements} which can be performed on the system yielding the \textit{outcomes} $A^i, B^j, ... $ ($i \! \in \! I, j\! \in \! J, ...$) respectively. Suppose that the measurements are repeatable and we perform them many times and obtain stable long-run relative frequencies for the outcomes in each state:
\begin{eqnarray*}
\frac{\#(A^i \w a \w s)}{\#(a \w s)} \quad , \quad  \frac{\#(B^j \w b \w r)}{\#(b \w r)}  \quad , \quad  \dots
\end{eqnarray*}
These relative frequencies allow us to introduce the \textit{conditional probabilities} of obtaining certain outcomes given that the system has been prepared in certain states and the appropriate measurements have been performed:
\begin{eqnarray*}
p(A^i|a \w s) \quad , \quad p(B^j|b \w r) \quad , \quad \dots
\end{eqnarray*}
We call a state $s \in S$ an \textit{eigenstate} of the measurement $a$ if 
\begin{eqnarray} \label{eigenstate}
p(A^i|a \w s) \in \{0,1\} \quad \quad \quad \mbox{for all} \, \, i \in I
\end{eqnarray}

If two measurements, say $a$ and $b$, can be \textit{jointly} or \textit{simultaneously} performed, then the joint frequencies
\begin{eqnarray*}
\frac{\#(A^i \w B^j \w a \w b \w s)}{\#(a \w b \w s)}
\end{eqnarray*}
are also well-defined which allows us to introduce the \textit{joint conditional probabilities}:
\begin{eqnarray*}
p(A^i \w B^j|a \w b \w s)
\end{eqnarray*}
Jointly or simultaneously performable measurements are also called \textit{comeasurable}.

Whether two measurements are comeasurable is a physical question. One can measure the width and the length of a table at the same time. But one cannot jointly check---using Arthur Fine's example---whether a given piece of wood is combustible \textit{and} whether it can float on water. The two measurements cannot be simultaneously performed; you cannot burn the piece of wood while in water. Similarly, you are not going to burn the piece of wood along with throwing it in water---unless you want to test whether the ash floats.

Let $M$ denote the set of \textit{all} measurements (basic and joint) physically performable on a system and let the variables $x, y$ range over the measurements in $M$. The outcomes of $x$ and $y$ are denoted by $X^k$ and $Y^l$, ($k \in K_x, l \in L_y$), respectively, and the set of outcomes of all measurements is denoted by $O = \cup_x \{X^k\}$. Similarly, let the variable $r$ range over the preparations $s,t, ... \in S$ of the system. An operational theory is then given by a set of conditional probabilities of the outcomes for the various basic and joint measurements in the various preparations:
\begin{eqnarray}\label{optheory}
p(X^k|x\w r) \quad \quad \quad \mbox{for all} \, \, k \in K_x, \, \, x \in M  \, \, \mbox{and} \, \, r \in S
\end{eqnarray}
which add up to 1 if we sum up for $k$.

Measurements which are \textit{not} jointly measurable are not to be conflated with \textit{disturbing} measurements. Consider the following example. In the army one performs two tests: shooting test ($a$) and tightrope walking ($b$). The two tests are jointly measurable; soldiers can well walk on a thin rope and shoot in the meanwhile. However, their performance in shooting  is heavily influenced by whether they are walking on a rope or not while shooting. Thus, two simultaneous measurements $a$ and $b$ are called \textit{non-disturbing} if 
\begin{eqnarray}
p(A^i|a \w b \w r) &=& p(A^i|a \w r) \label{nondist1} \quad \quad \quad \mbox{for all} \, \, i \in I  \, \, \mbox{and} \, \, r \in S \\
p(B^j|a \w b \w r) &=& p(B^j|b \w r) \label{nondist2} \quad \quad \quad \mbox{for all} \, \, j \in J  \, \, \mbox{and} \, \, r \in S
\end{eqnarray}
For spacelike separated measurements no-disturbance is equivalent to \textit{no-signaling}.

A non-disturbing operational theory can be characterized in the following compact way. First note that there is a natural partial ordering on the measurements of an operational theory which expresses ``how joint'' the measurements are. $a \w b$ is ``more joint'' than $a$ or $b$. Call the set of basic measurements $\{a, b, ...\}$ the \textit{basis} of a measurement $x$, if $x = a \w b \w ...$. Now, for two measurements $x,y \in M$ let $x \geqslant y$ if the basis of $x$ is contained in or equal to the basis of $y$. Using this partial ordering, an operational theory is non-disturbing if:
\begin{eqnarray}\label{non-dist-theor}
p(X^k|x \w r) = p(X^k|y \w r) \quad \quad \quad \mbox{for all} \, \, k \in K_x, \, \, r \in S \, \, \mbox{and} \, \, x,y \in M  \, \, \mbox{such that} \, \, x \geqslant y 
\end{eqnarray}

Denote by $M^m$ the set of \textit{maximally joint measurements}, that is, the set of measurements $x$ for which there is no other measurement $y$ such that $x\geqslant y$. For a non-disturbing operational theory it is enough to specify the conditional probabilities (\ref{optheory}) for all $x \in M^m$; all other conditional probabilities will then be set by (\ref{non-dist-theor}).

\section{Ontological models}\label{Sec:Ont}

The role of an \textit{ontological model}\footnote{\textit{Cf.} Spekkens (2005).} (hidden variable model) is to account for the conditional probabilities of an operational theory in terms of underlying realistic entities called \textit{ontic states} (hidden variables, elements of reality, beables). An ontological model defines the preparations of the system in terms of \textit{distributions} over the ontic states and specifies the response of the system to the different measurements in the different ontic states in terms of the so-called \textit{response functions}. The ontological model is successful if the conditional probabilities of the operational theory can be recovered in terms of these distributions and response functions.

Mathematically, the provision of an ontological model starts with the specification the set $\Lambda$ of ontic states and a variable $\l$ running over $\Lambda$. To make things simple we assume that $\Lambda$ is countable.\footnote{But nothing hinges on the cardinality of $\Lambda$.} Next, we associate with each preparation a \textit{probability distribution} over the ontic states:
\begin{eqnarray}\label{HVM}
p(\l|r)  \quad \quad \quad \mbox{for all} \, \, r \in S
\end{eqnarray}
and to each measurement and ontic state a set of \textit{response functions} that is, a set of conditional probabilities
\begin{eqnarray}\label{HVM2}
p(X^k|x\w \l) \quad \quad \quad \mbox{for all} \, \, k \in K_x, \, \, x \in M  \, \, \mbox{and} \, \, \l \in \Lambda
\end{eqnarray}
again with the obvious normalization.

One can also impose two natural screening-off conditions expressing the independence of the preparations, measurements and ontic states. The first screening-off condition, called \textit{no-conspiracy}, requires that the probability distributions do not depend causally, and hence probabilistically, on the measurements performed on the system:
\begin{eqnarray}\label{nocons}
p(\l|r) = p(\l| r \w x )  \quad \quad \quad \mbox{for all} \, \, x \in M \, \, \mbox{and} \, \, r \in S
\end{eqnarray}
The second screening-off condition, called \textit{$\l$-sufficiency}, requires that the response functions do not depend on the preparations in which the ontic states are featuring: 
\begin{eqnarray}\label{lsuff}
p(X^k|x\w \l) = p(X^k|x\w \l \w r ) \quad \quad \quad \mbox{for all} \, \, k \in K_x, \, \, x \in M,  \, \,  \l \in \Lambda \, \, \mbox{and} \, \, r \in S
\end{eqnarray}
By means of (\ref{nocons})-(\ref{lsuff}) and using the theorem of total probability one obtains:
\begin{eqnarray}  \label{recov}
p(X^k|x\w r) &=& \sum_\l p(X^k|x\w \l \w r)  \, p(\l|r \w x) \nonumber \\
&=& \sum_\l p(X^k|x\w \l)  \, p(\l|r)  \quad \quad \quad \mbox{for all} \, \, k \in K_x, \, \, x \in M  \, \, \mbox{and} \, \, r \in S
\end{eqnarray}
That is, one recovers the operational theory from the ontological model in terms of the probability distributions and response functions.

An ontic state $\l$ with respect to a measurement $x$ is called \textit{value-definite} if
\begin{eqnarray}\label{det}
p(X^k|x\w \l) \in \{0,1\}\quad \quad \quad \mbox{for all} \, \, k \in K_x
\end{eqnarray}
otherwise it is called \textit{probabilistic}. Recall that one and same $\l$ can be value-definite for the one measurement and probabilistic for the other. An ontological model is called \textit{value-definite} if (\ref{det}) holds for all $x \in M^m$; otherwise it is called \textit{probabilistic}.

\section{Noncontextuality}\label{Sec:NC}

Ontological models, both value-definite and probabilistic, trivially exist for an operational theory if no further constraints are put on them. But now require that the ontological model is noncontextual.

\noi
\textit{An ontological model is (simultaneous) noncontextual if every ontic state determines the probability of the outcomes of every measurement independently of what other measurements are simultaneously performed; otherwise is contextual.} 

\noi
(Simultaneous) noncontextuality can be formally expressed as follows:
\begin{eqnarray}\label{mynoncontext}
p(X^k|x \w \l) = p(X^k|y \w \l) \quad \quad \quad \mbox{for all} \, \, k \in K_x, \, \, \l \in \Lambda \, \, \mbox{and} \, \, x,y \in M  \, \, \mbox{such that} \, \, x \geqslant y 
\end{eqnarray}
In other words, each ontic state uniquely determines the probability of all outcomes of a given measurement irrespective of what other measurements are co-measured. A specific consequence of (\ref{mynoncontext}) is that the conditional probabilities of all basic measurements will be fixed irrespective of what other measurements they are co-measured with.

Observe, that noncontextuality\footnote{From now on, I drop the qualifier ``simultaneous'' but the term ``noncontextuality'' will continue to mean ``simultaneous noncontextuality'' as defined in (\ref{mynoncontext}).} (\ref{mynoncontext}) is almost the same requirement as no-disturbance (\ref{non-dist-theor}), except that the latter is required for the preparations while the former is required for the ontic states. Consequently, noncontextuality provides a neat explanation for why an operational theory is non-disturbing: if an ontological model for an operational theory satisfies noncontextuality (\ref{mynoncontext}) (and also no-conspiracy (\ref{nocons}) and $\l$-sufficiency (\ref{lsuff})), then the operational theory will satisfy no-disturbance (\ref{non-dist-theor}). Hence, the assumption of noncontextuality is a kind of inference to the best explanation for the non-disturbing character of an operational theory.

Some notes are in place here. (i) Noncontextuality (\ref{mynoncontext}) is a generalization of Shimony's (1986) \textit{parameter independence} for situations when the simultaneous measurements are not necessarily spacelike separated. 

(ii) If a \textit{value-definite} ontological model is noncontextual, then (\ref{det}) will hold for all $x \in M$ (and not just for $x \in M^m$). 

(iii) Noncontextuality of an ontological model does not generally imply \textit{factorization}:
\begin{eqnarray}\label{factor}
p(X^k \w Y^l|x \w y \w \l) = p(X^k|x \w \l) \, p(Y^l|y \w \l)  &\quad  \quad & \mbox{for all} \, \, k \in K_x, \, \, l \in L_y, \, \, \l \in \Lambda \nonumber \\
 &\quad \quad  & \mbox{and} \, \, x,y, x\w y \in M
\end{eqnarray}
But it does if the ontological model is value-definite.

(iv) Noncontextuality as defined in (\ref{mynoncontext}) resembles to the concept of noncontextuality of Simon et al.\! (2001) but differs from that of Spekkens (2005) and other operationalists. Below I will refer to this latter concept as ``Spekkens' condition.''

\section{Quantum mechanical representation}\label{Sec:QM}

On the minimal interpretation QM is an operational theory which provides conditional probabilities for the outcomes of different measurements in different states. Thus, the empirical content of QM could be expressed simply by listing the various conditional probabilities. However, in the standard formalism these conditional probabilities get represented in a linear algebraic fashion. The physical system is associated with a Hilbert space; each state $r \in S$ is represented by a density operator $\hat{\boldsymbol{\rho}}_r$; each measurement $x \in M$ by a self-adjoint operator $\hat{\textbf{x}}$; and the outcome $X^k$ of $x$ by the orthogonal spectral projection $\hat{\textbf{P}}_{\!x}^k$ of $\hat{\textbf{x}}$ with eigenvalue $X^k$. The representation is connected to experience by the Born rule:  
\begin{eqnarray}\label{born}
\mbox{Tr}(\hat{\boldsymbol{\rho}}_r\hat{\textbf{P}}_x^k) = p(X^k|x \w r) \quad \quad \quad \mbox{for all} \, \, k \in K_x, \, \, x \in M  \, \, \mbox{and} \, \, r \in S
\end{eqnarray}
where Tr is the trace function.

Now, if $a$ and $b$ are comeasurable, then $a \w b$ gets represented in QM by commuting operators $\hat{\textbf{a}}$ and $\hat{\textbf{b}}$. But if $\hat{\textbf{a}}$ and $\hat{\textbf{b}}$ are commuting, then $a$ and $b$ will turn out to be non-disturbing: 
\begin{eqnarray*}
p(A^i|a \w b \w r) = \sum_j p(A^i \w B^j |a \w b \w r) =&& \nonumber \\
 \sum_j \mbox{Tr}(\hat{\boldsymbol{\rho}}_r \hat{\textbf{P}}_a^i\hat{\textbf{P}}_b^j) = \mbox{Tr}(\hat{\boldsymbol{\rho}}_r \hat{\textbf{P}}_a^i) = p(A^i|a \w r) \label{qnondist1} && \quad \quad \quad \mbox{for all} \, \, i \in I  \, \, \mbox{and} \, \, r \in S 
\end{eqnarray*}
and similarly for $p(B^j|a \w b \w r)$. Thus, the quantum mechanical representation of joint measurements implies that QM cannot represent comeasurable but disturbing measurements. In other words, only non-disturbing operational theories can have a quantum mechanical representation.

Being an operational theory, one can search for an ontological model for QM. The KS arguments are intending to rule out such an ontological model if it is both value-definite and noncontextual.\footnote{The restriction to value-definiteness is dropped in certain arguments (Mazurek et al. 2016), but here noncontextuality is defined as \textit{measurement} noncontextuality \textit{á la} Spekkens (2005) and not as (\ref{mynoncontext}).} In the following sections I pick a special KS theorem, the Peres-Mermin square (Peres, 1990; Mermin, 1992) and investigate whether it can be given a unique realization, that is, an operational theory composed of 9 simultaneous measurements which does not admit a value-definite, noncontextual ontological model.

\section{An example: the Peres-Mermin square}\label{Sec:PM}

Consider the following $3 \! \times \! 3$ matrix of self-adjoint operators:
\begin{center}
\begin{tabular}{cccc}
& \quad $\hat{\boldsymbol{a}} \equiv \hat{\boldsymbol{\sigma}}_3 \otimes \hat{\UN}$ \quad &  \quad $\hat{\boldsymbol{b}} \equiv \hat{\UN} \otimes \hat{\boldsymbol{\sigma}}_3$  \quad &  \quad $\hat{\boldsymbol{c}} \equiv \hat{\boldsymbol{\sigma}}_3 \otimes \hat{\boldsymbol{\sigma}}_3$  \quad \\ 
& & & \\ 
& \quad $\hat{\boldsymbol{d}} \equiv \hat{\UN} \otimes \hat{\boldsymbol{\sigma}}_1$  \quad &  \quad $\hat{\boldsymbol{e}} \equiv \hat{\boldsymbol{\sigma}}_1 \otimes  \hat{\UN}$  \quad &  \quad $\hat{\boldsymbol{f}} \equiv \hat{\boldsymbol{\sigma}}_1 \otimes \hat{\boldsymbol{\sigma}}_1$  \quad \\ 
& & & \\ 
& \quad $\hat{\boldsymbol{g}} \equiv \hat{\boldsymbol{\sigma}}_3 \otimes \hat{\boldsymbol{\sigma}}_1$  \quad &  \quad $\hat{\boldsymbol{h}} \equiv \hat{\boldsymbol{\sigma}}_1 \otimes \hat{\boldsymbol{\sigma}}_3$  \quad &  \quad $\hat{\boldsymbol{i}} \equiv \hat{\boldsymbol{\sigma}}_2 \otimes \hat{\boldsymbol{\sigma}}_2$  \quad 
\end{tabular}
\end{center}
where $\hat{\boldsymbol{\sigma}}_1, \hat{\boldsymbol{\sigma}}_2$ and $\hat{\boldsymbol{\sigma}}_3$ are the Pauli operators and $\hat{\UN}$ is the unit operator on the two dimensional complex Hilbert space. The operators in the matrix are arranged in such a way that two operators are commuting if and only if they are in the same row or in the same column. Each operator in the matrix has two eigenvalues, $\pm1$. Denote the spectral projections of the operators $\hat{\boldsymbol{a}}, \hat{\boldsymbol{b}}, \hat{\boldsymbol{c}}, ...$ associated with the eigenvalues $\pm1$ by $\hat{\textbf{P}}_a^\pm, \hat{\textbf{P}}_b^\pm, \hat{\textbf{P}}_c^\pm, ...$, respectively. Let the variables $\hat{\boldsymbol{x}}, \hat{\boldsymbol{y}}$, and $\hat{\boldsymbol{z}}$ range over the operators of the Peres-Mermin square. Denote the spectral projections of $\hat{\boldsymbol{x}}, \hat{\boldsymbol{y}}$, and $\hat{\boldsymbol{z}}$ by $\hat{\textbf{P}}_x^j$, $\hat{\textbf{P}}_y^k$, and $\hat{\textbf{P}}_z^l$ $(j, k, l = \pm1$), respectively. The set of states $S$ is represented by the set of density operators on the two dimensional complex Hilbert space (which also include the common eigenstates for each subset of mutually commuting operators).

The quantum probabilities for the spectral projections of the three vertical and three horizontal commuting triples of operators are given by the trace formula:
\begin{eqnarray}\label{qmermin}
\mbox{Tr}(\hat{\boldsymbol{\rho}}_r\hat{\textbf{P}}_x^\pm \hat{\textbf{P}}_y^\pm \hat{\textbf{P}}_z^\pm)   \quad \quad \quad \mbox{for all} \, \, \rho_r \, \, \mbox{density operators}
\end{eqnarray}
Now, it turns out that these quantum probabilities are non-zero only for certain combinations of spectral projections for a given commuting triple (irrespective of the quantum state). More specifically, for the third vertical triple $(\{\hat{\boldsymbol{c}}, \hat{\boldsymbol{f}},\hat{\boldsymbol{i}}\})$ the quantum probabilities are non-zero only for those combinations of projections for which the product of the associated eigenvalues is $-1$. For the other five triples this product must be $+1$. That is,
\begin{eqnarray}\label{qmermin2}
\mbox{Tr}(\hat{\boldsymbol{\rho}}_r\hat{\textbf{P}}_x^j \hat{\textbf{P}}_y^k \hat{\textbf{P}}_z^l)  \neq 0 \, \, \, \mbox{only if} \, \, \left\{ \begin{array}{ll} j\cdot k \cdot l = -1 & \mbox{if} \, \, \{\hat{\boldsymbol{x}}, \hat{\boldsymbol{y}},\hat{\boldsymbol{z}}\} = \{\hat{\boldsymbol{c}}, \hat{\boldsymbol{f}},\hat{\boldsymbol{i}}\} \\ j\cdot k \cdot l   = +1 & \mbox{otherwise} \end{array} \right.
\end{eqnarray}
Note that these admissible combinations of eigenvalues are also associated with the four common eigenstates of the triplet in question. 

Now, these admissible combinations of eigenvalues provide a constraint on the value assignments that is, on the functions sending each of the nine operators of the Peres-Mermin square to one of their eigenvalues, that is, to $\pm1$. The constraint is that the product of the numbers in each row and column should be $+1$, except for the third column where it should be $-1$. It is easy to see that no such value assignment exists.

But does this no-go result prove that QM does not admit a noncontextual value-definite ontological model? Not until the Peres-Mermin square is given a unique physical realization.

\section{An operational theory realizing the Peres-Mermin square}\label{Sec:OTPM}

Consider an operational theory with 9 basic measurements:
\begin{center}
\begin{tabular}{cccc}
& \quad $a$ \quad &  \quad $b$  \quad &  \quad $c$  \quad \\ 
& & & \\ 
& \quad $d$  \quad &  \quad $e$  \quad &  \quad $f$  \quad \\ 
& & & \\ 
& \quad $g$  \quad &  \quad $h$  \quad &  \quad $i$  \quad 
\end{tabular}
\end{center}
The $3 \times 3$ matrix in which the measurements are arranged is to express now \textit{comeasurability relations}: measurements are simultaneously measurable if and only if they are in the same row or in the same column. 

Each measurement can have two outcomes, $A^\pm, B^\pm,C^\pm, ... = \pm1$. Let the variables $x, y$ and $z$ range over the basic measurements $M^b$. Denote the outcomes of $x, y$ and $z$ by $X^j, Y^k$ and $Z^l$ $(j, k, l = \pm1$), respectively. Let the conditional probability of the 6 different maximally joint measurements be: 
\begin{eqnarray}\label{mermin}
p(X^\pm \w Y^\pm \w Z^\pm\, |\,x \w y \w z \w r)   \quad \quad \quad \mbox{for all} \, \, r \in S
\end{eqnarray}
Suppose furthermore that the condition probabilities of all other non-maximally joint measurements can be obtained from (\ref{mermin}) by marginalization. Thus, (\ref{mermin}) characterizes a non-disturbing operational theory.

Now, suppose that the operational theory (\ref{mermin}) is a physical realization of the Peres-Mermin square in the sense that the quantum probabilities (\ref{qmermin}) in the Peres-Mermin square represent just the conditional probabilities (\ref{mermin}) via the Born rule (\ref{born}). That is,
\begin{eqnarray}\label{qmermin3}
\mbox{Tr}(\hat{\boldsymbol{\rho}}_r\hat{\textbf{P}}_x^j \hat{\textbf{P}}_y^k \hat{\textbf{P}}_z^l)  = p(X^j \w Y^k \w Z^l|x \w y \w z \w r)   \quad \quad \quad \mbox{for all} \, \, r \in S
\end{eqnarray}
Note that (\ref{qmermin3}) is well-defined since the operators on the left hand side are mutually commuting if and only if the represented measurements on the right hand side are comeasurable.  Also note that the operational theory (\ref{mermin}) is a \textit{unique realization} of the Peres-Mermin square, since every operator is associated with a different measurement. As we saw in the Introduction, only unique realizations can decide on the status of noncontextuality in QM. (In Section \ref{Sec:KS(iii)} we will see what non-unique realizations can do.)

From (\ref{qmermin2}) and (\ref{qmermin3}) it follows that the support of the probability distributions over the outcomes that is, the set of possible outcomes for each maximally joint measurement $x\w y \w z$ and each preparation $r \in S$ is the following:
\begin{eqnarray}\label{mermin2}
p(X^j \w Y^k \w Z^l \, |\,x \w y \w z \w r)  \neq 0 \, \, \, \mbox{only if} \, \, \left\{ \begin{array}{ll} j\cdot k \cdot l = -1 & \mbox{if} \, \, \{x,y,z\} = \{c,f,i\} \\ j\cdot k \cdot l   = +1 & \mbox{otherwise} \end{array} \right.
\end{eqnarray}
that is, the conditional probability is non-zero only for such joint outcomes which contain an odd number of $+1$s and an even number of $-1$s in each row and column, except for the last column where the number of $+1$s is even and the number of $-1$s is odd.

Does the operational theory (\ref{mermin}) have a noncontextual value-definite ontological model?

Assume (contrary to fact) that there is such a model with response functions:\footnote{Note that for this argument we don't need the probability distributions $p(\l|r)$.}
\begin{eqnarray}\label{mermin3}
p(X^\pm \w Y^\pm \w Z^\pm \, |\,x \w y \w z \w \l)   \quad \quad \quad \mbox{for all} \, \, \l \in \Lambda
\end{eqnarray}
Being noncontextual and value-definite, the response functions are factorizing: 
\begin{eqnarray}\label{mermin3}
p(X^\pm \w Y^\pm \w Z^\pm \, |\,x \w y \w z \w \l) = p(X^\pm  \,|\,x \w \l) \, p(Y^\pm  \, |\, y \w \l) \, p(Z^\pm \, |\, z \w \l)   
\end{eqnarray}
for all $\l \in \Lambda$. Thus, the ontological model can be characterized by the extremal conditional probabilities:
\begin{eqnarray}\label{mermin4}
p(X^\pm \,|\,x \w \l) \in \{0,1\} \quad \quad \quad \mbox{for all} \, \, x \in M^b \, \, \, \mbox{and} \, \,\l \in \Lambda
\end{eqnarray}

However, the support (\ref{mermin2}) of the operational theory restricts the possible extremal conditional probabilities. Namely, for any three simultaneous measurements $x$, $y$ and $z$ in $M^b$ and $\l \in \Lambda$ one requires that
\begin{eqnarray}\label{mermin5}
p(X^j \, |\,x \w \l) \, p(Y^k \, |\, y \w \l) \, p(Z^l \, |\, z \w \l)  = 1 \, \, \, \mbox{only if} \, \, \left\{ \begin{array}{ll} j\cdot k \cdot l = -1 & \mbox{if} \, \, \{x,y,z\} = \{c,f,i\} \\ j\cdot k \cdot l  = +1 & \mbox{otherwise} \end{array} \right.
\end{eqnarray}
otherwise there could be some ontic states which, if prepared (that is, $p(\l|r) \neq 0$ for some $r \in S$), would render at least one conditional probability in (\ref{mermin}) non-zero outside the support (\ref{mermin2}).

However, it is easy to see that there is no such a set of conditional probabilities (\ref{mermin4}) which satisfies (\ref{mermin5}). This is due to the impossibility to fill in a  $3 \times 3$ matrix with $\pm 1$s such that the product of the numbers in each row and column is $+1$, except for the last column where it is $-1$. Consequently, the operational theory (\ref{mermin}) does not have a noncontextual value-definite ontological model.

\noi
Let me briefly reflect on the question of experimental testability of the above operational theory. Suppose that in a real experiment the support equation (\ref{mermin2}) cannot be sharply validated but only up to a fraction $1-\epsilon$ of all runs. How small $\epsilon$ should be so that a noncontextual value-definite ontological model for the operational theory can still be ruled out?

Suppose \textit{a contrario} that the ontological model is noncontextual and it conforms to the measurement statistics as much as possible, that is, for all $\l \in \Lambda$ only one of the six constraints (\ref{mermin5}) is violated. (For example some $\l$ assigns $+1$ to all 9 measurements, violating thus the constraint of the third column but respecting all the other five, etc.) Since there are six different triply joint measurements (of the three rows and three columns), hence---modulo some conspiracy---there is a 1/6 probability for any $\l$ that a certain joint measurement will pick just that triple for which (\ref{mermin5}) is violated. Since each such measurement will contribute to the violation of (\ref{mermin2}), (\ref{mermin2}) will be violated in a fraction of 1/6 of all runs. Consequently, if in a real experiment $\epsilon$ is smaller than 1/6, then the experiment will rule out a noncontextual value-definite ontological model for the operational theory. 

This argument is a special case of a general argument provided by Simon et al. (2001) and Larsson (2002) in the defense of the KS arguments against the so-called \textit{finite precision loophole} argument of Meyer (1999), and Clifton and Kent (2000). As Barrett and Kent (2004, Section 4.3) nicely point out, the finite precision loophole is effective only if noncontextuality is defined in terms of operators on a Hilbert space and not operationally in terms of measurements---in short, only if KS arguments are understood as KS \textit{theorems}. Thus, the finite precision loophole arguments do not nullify the KS arguments based on the above operational theory.

\section{Do spin measurements realize the Peres-Mermin square?}\label{Sec:SMPM}

The only question that remains is thus whether there exists an operational theory physically realizing the Peres-Mermin square?

The first idea that comes to mind is the standard spin measurements. Suppose that the operator $\hat{\boldsymbol{\sigma}}_i \otimes \hat{\boldsymbol{\sigma}}_j$ ($i,j= 1,2,3$) represents the measurement that first we perform two spin measurements by two Stern-Gerlach magnets on a pair of spin-$\frac{1}{2}$ particles in directions $\vec{i}$ and $\vec{j}$, respectively ($\vec{i}, \vec{j} \in \{\vec{x}, \vec{y},\vec{z}\}$; $\vec{x}, \vec{y}$ and $\vec{z}$ are mutually perpendicular); and second we check whether the outcomes of the measurements on the opposite wings are the same ($+1$) or not ($-1$). Denote this composite measurement, symbolically, by $(s_i \w s_j)^\pm$. Furthermore, let $\hat{\boldsymbol{\sigma}}_i \otimes \hat{\UN}$ ($i= 1,2,3$) and $\hat{\UN} \otimes \hat{\boldsymbol{\sigma}}_j$ ($j= 1,2,3$) represent that we perform the spin measurement only on the left and right particle, respectively. Denote these singular spin measurements, symbolically, by $s_i \w 1$ and $1 \w s_j$, respectively. Then, the measurements realizing uniquely the Peres-Mermin square read as follows:
\begin{center}
\begin{tabular}{lll}
$a \equiv s_3 \w 1$ \quad &  \quad $b \equiv 1 \w s_3$  \quad &  \quad $c \equiv (s_3 \w s_3)^\pm$ \\ 
& & \\ 
$d \equiv 1 \w s_1$  \quad &  \quad $e \equiv s_1 \w  1$  \quad &  \quad $f \equiv (s_1 \w s_1)^\pm$ \\ 
& & \\ 
$g \equiv (s_3 \w s_1)^\pm$  \quad &  \quad $h \equiv (s_1 \w s_3)^\pm$  \quad &  \quad $i \equiv (s_2 \w s_2)^\pm$ 
\end{tabular}
\end{center}

Unfortunately, however, only four of the six commuting subsets of operators represent simultaneous measurements: the first two rows and the first two columns. Measurements in the third row and in the third column are, however, \textit{not} comeasurable. For example, the measurements $c$, $f$ and $i$ in the third column, that is, the spin measurements in directions $\vec{z}\!-\!\vec{z}$, $\vec{x}\!-\!\vec{x}$, and $\vec{y}\!-\!\vec{y}$ cannot be simultaneously performed: one cannot turn the Stern-Gerlach magnets in directions $\vec{z}\!-\!\vec{z}$, $\vec{x}\!-\!\vec{x}$, and $\vec{y}\!-\!\vec{y}$ at the same time. Consequently, although the left hand side of (\ref{qmermin3}) exists, the right hand side is ill-defined for the third column and also for the third row. The quantum probabilities
\begin{eqnarray*}
\mbox{Tr}(\hat{\boldsymbol{\rho}}_r \, \hat{\textbf{P}}_c^\pm \, \hat{\textbf{P}}_f^\pm \, \hat{\textbf{P}}_i^\pm) \\
\mbox{Tr}(\hat{\boldsymbol{\rho}}_r \, \hat{\textbf{P}}_g^\pm \, \hat{\textbf{P}}_h^\pm \, \hat{\textbf{P}}_i^\pm)
\end{eqnarray*}
cannot be interpreted as conditional probabilities 
\begin{eqnarray*}
p(C^\pm \w F^\pm \w I^\pm\, |\,c \w f \w i \w r)   \\
p(G^\pm \w H^\pm \w I^\pm\, |\,g \w h \w i \w r)  
\end{eqnarray*}
and hence neither their support is defined. So one does not have the constraint 
\begin{eqnarray}
p(C^i \,|\, c \w \l) \, p(F^j \,|\, f \w \l) \, p(I^k \,|\, i \w \l) = 1 & \quad \quad \quad & \mbox{only if}  \, \, \, j\cdot k \cdot l  = -1 \label{needed1} \\
p(G^i \,|\, c \w \l) \, p(H^j \,|\, f \w \l) \, p(I^k \,|\, i \w \l) = 1 & \quad \quad \quad & \mbox{only if}  \, \, \, j\cdot k \cdot l  = 1 \label{needed2}
\end{eqnarray}
for the ontic states in the third column and third row and hence cannot arrive at the contradiction outlined above. The whole argumentation collapses. In short, the standard spin measurement does not realize the Peres-Mermin square in form of an operational theory (\ref{mermin}), and consequently does not provide a physical realization for a quantum mechanical scenario for which a noncontextual value-definite ontological model could be ruled out.

Obviously, the standard realization of the above operators in terms of spin measurements is not the only possible physical realization. One may well come up with another unique realization on which the measurements are comeasurable if and only if the representing operators are commuting. However, I know of no such realization. And the burden of proof is on those who claim that the above arrangement of operators exclude a noncontextual value-definite ontological model for QM. An uninterpreted formalism cannot prove anything about the outer world.\footnote{But one might respond: why not to measure $c$, $f$ and $i$ simultaneously by one single ``global'' measurement (Reck et al., 1994)? We return to this question in Section \ref{Sec:KS(iii)}.}

\noi
Perhaps it is worth reflecting for a moment on the relation of commutativity and comeasurability (see Park and Margenau, 1968). Comeasurability is used in two different meanings in quantum physics. First, two measurements are called comeasurable (compatible, simultaneously measurable) if, performing them one \textit{after} another, the first measurement does not alter the outcome statistics of the second one. Obviously, this usage of the term ``simultaneous'' is metaphoric and has no bearing on the KS arguments. 

The other meaning is the one we use throughout this paper: two measurement are comeasurable if they can physically be performed at the same time on the same system. Note, however, that this notion of comeasurability and the notion of commutativity are not synonym expressions. From the simple fact that two measurements are represented by commuting operators it does \textit{not} follow that the measurements are simultaneously performable. Comeasurability is a physical question which cannot be simply read off from their representation. Simultaneous measurements get represented in QM by commuting operators. But the converse is not true. Not \textit{all} commuting operators represent simultaneous measurements. Consider the following three pairs of commuting operators:
\begin{eqnarray*}
\left[\hat{\boldsymbol{S}}_1^2 \, ,\, \hat{\boldsymbol{S}}_2^2 \right] &=& 0 \\
\left[\hat{\boldsymbol{\sigma}}_1 \otimes \hat{\boldsymbol{\sigma}}_3 \, ,\, \hat{\boldsymbol{\sigma}}_3 \otimes \hat{\boldsymbol{\sigma}}_1 \right] &=& 0 \\
\left[\hat{\boldsymbol{\sigma}}_1 \otimes \hat{\boldsymbol{\sigma}}_1 \otimes \hat{\boldsymbol{\sigma}}_1 \, ,\, \hat{\boldsymbol{\sigma}}_2 \otimes \hat{\boldsymbol{\sigma}}_2 \otimes \hat{\boldsymbol{\sigma}}_1 \right] &=& 0
\end{eqnarray*}
where $\hat{\boldsymbol{S}}_1, \hat{\boldsymbol{S}}_2$ and $\hat{\boldsymbol{\sigma}}_1, \hat{\boldsymbol{\sigma}}_2$ are spin-1 and spin-$\frac{1}{2}$ operators, respectively. Each pair is featuring in one or other of a renowned KS argument: the first pair in the original Kochen-Specker (1967) argument; the second in Peres' (1990) and Mermin's (1992) version and also in Cabello's (1997) version; and the third in the GHZ (1989) version of the argument. However, none of them can be interpreted as operators representing simultaneous spin measurements on pairs or triples of spin-$1$ or spin-$\frac{1}{2}$ particles. But in the absence of a unique realization of a KS graph where commuting operators represent simultaneous measurements, the no-go results do not prove that QM does not admit a noncontextual value-definite ontological model.

How then the above KS arguments work?

\section{Three types of Kochen-Specker arguments}\label{Sec:Three} 

To see the problem more clearly, it is worth introducing the following categorization. Suppose we are given a unique realization, that is, a KS graph \textit{and} an associated operational theory realizing the operators on the graph in a one-to-one manner. Now, one can cast the KS arguments into \textit{three} types according to the number of subsets of mutually commuting operators (operators on a hyperedge) which do \textit{not} represent simultaneous measurements in the associated operational theory:
\begin{description}
 \item[Arguments of type I: ]  where \textit{all} commuting subsets represent simultaneous measurements;
 \item[Arguments of type II: ] where \textit{all but one} commuting subset represent simultaneous measurements;
 \item[Arguments of type III: ] where there is \textit{more than one} commuting subset \textit{not} representing simultaneous measurements.
\end{description}

As it will turn out soon, there is a huge difference in the efficacy of the three types of arguments. 

It is only KS arguments of type I which provide a state-independent (algebraic) proof for quantum contextuality, since for these arguments FUNC can be physically justified by the probability distribution of the joint outcomes of simultaneous measurements. Unfortunately, I am not aware of any argument of type I. In other words, I am not aware of any unique realization of any KS graph where all commuting subsets of operators would represent simultaneous measurements. Consequently, I am also not aware of any state-independent argument proving quantum contextuality.

KS arguments of type II do exist but they provide only a state-dependent proof for quantum contextuality. An example for such an arguments is the GHZ argument. I return to this argument in the next section.  

Finally, KS arguments of type III abound. The Peres-Mermin square with the standard spin realization is one example: the number of commuting subsets not representing simultaneous measurements is two, the three operators in the third row and the three operators in the third column. Another example for arguments of type III is the original KS graph with 117 vertices with the standard spin realization. Here \textit{none} of the commuting subsets represents simultaneous measurements since the spin measurements for three orthogonal directions cannot be simultaneously performed. In section \ref{Sec:KS(iii)}, I will argue that arguments of type III are inconclusive in proving quantum contextuality. To get a contradiction, they need to flip to a non-unique (hyperedge-based) realization and invoke Spekkens' condition. However, by abandoning Spekkens' condition the contradiction can be avoided.

\section{Kochen-Specker arguments of type II}\label{Sec:KS(ii)}

Let us see first the KS arguments of type II. A prototype of such  arguments is the GHZ argument. The GHZ graph (pentagram) reads as follows:
{\small
\begin{center}
\begin{tabular}{llrr}
\multicolumn{4}{c}{$\hat{\boldsymbol{\sigma}}_2 \otimes \hat{\UN} \otimes \hat{\UN}$}\\
& & & \\ 
& & & \\ 
$\hat{\boldsymbol{\sigma}}_1 \otimes \hat{\boldsymbol{\sigma}}_1 \otimes \hat{\boldsymbol{\sigma}}_1$   &
$\hat{\boldsymbol{\sigma}}_2 \otimes \hat{\boldsymbol{\sigma}}_2 \otimes \hat{\boldsymbol{\sigma}}_1$    &
$\hat{\boldsymbol{\sigma}}_2 \otimes \hat{\boldsymbol{\sigma}}_1 \otimes \hat{\boldsymbol{\sigma}}_2$   &
$\hat{\boldsymbol{\sigma}}_1 \otimes \hat{\boldsymbol{\sigma}}_2 \otimes \hat{\boldsymbol{\sigma}}_2$  \\ 
& & & \\ 
& & & \\ 
& $\hat{\UN} \otimes  \hat{\UN} \otimes \hat{\boldsymbol{\sigma}}_1$  & $\hat{\UN} \otimes \hat{\UN} \otimes \hat{\boldsymbol{\sigma}}_2$  &  \\
& & & \\ 
& & & \\ 
\multicolumn{4}{c}{$\hat{\boldsymbol{\sigma}}_1 \otimes \hat{\UN} \otimes  \hat{\UN}$}\\
& & & \\ 
& & & \\ 
\multicolumn{2}{c}{$\hat{\UN} \otimes \hat{\boldsymbol{\sigma}}_2 \otimes  \hat{\UN}$} & \multicolumn{2}{c}{$\hat{\UN} \otimes \hat{\boldsymbol{\sigma}}_1 \otimes  \hat{\UN}$}
\end{tabular}
\end{center}}
On the standard spin realization of the GHZ graph, \textit{all but one} subsets of the mutually commuting operators can be interpreted as representing simultaneous measurements. Measurements represented by commuting operators on four of the five edges of the GHZ pentagram are comeasurable since they are performed on three spacelike separated subsystems. But the measurements represented by the operators on the fifth, horizontal edge are not comeasurable. 

How does then the KS argument work in the GHZ case?

The trick to circumvent the problem of non-comeasurability is to prepare the system in one of the common eigenstates of the measurements on the horizontal edge.\footnote{See (\ref{eigenstate}) for how an eigenstate for a \textit{measurement} is defined.} The outcome for each measurement on the horizontal edge will then be fixed even if the measurements are not comeasurable. The product of the possible outcomes of the four different measurements will turn out to be $-1$ in each common eigenstate. Now, the measurements on the other four lines of the GHZ pentagram are comeasurable, and the product of their possible joint outcomes in \textit{all} states (among them in the above common eigenstates) will be $+1$. This means that each ontic state in the support of these common eigenstates needs to assigns $\pm1$ to the individual measurements such that the product of these numbers is $+1$ in each line, except in the horizontal line where it is $-1$. Such value assignment, however, is impossible, which rules out a noncontextual value-definite ontological model for the GHZ scenario.

More generally, KS arguments of type II where all but one set of commuting operators represent simultaneous measurements are all state-dependent arguments. One needs to prepare the system in one of the common eigenstates of the non-comeasurable measurements to ``compensate'' the failure of comeasurability of these measurements. By doing so one obtains the same constraint on the response functions (necessary for deriving the contradiction) as one would obtain if the measurements were comeasurable. But note that these argument of type II cannot be transformed into a state-independent argument. They work only if the system is prepared in one of the common eigenstates of the operators representing non-comeasurable measurements.

\section{Kochen-Specker arguments of type III}\label{Sec:KS(iii)}

Finally, let us turn to the KS arguments of type III that is, to arguments where there is more than one commuting subset not representing simultaneous measurements. Here the strategy outlined in the previous section does not work. Even if one prepares the system in a common eigenstate of a set of operators representing non-comeasurable measurements, there remains at least one other set of non-comeasurable measurements for which the joint outcomes are not known. This blocks the KS argument since the constraint on the ontic state coming from this very set of measurements will be missing. 

One might however raise the question: Why not simply replace a commuting subset not representing simultaneous measurements by \textit{one single} measurement and apply certain functions on the result? Then the comeasurability problem would be solved.
 
Well, it is indeed a mathematical fact that for any finite set $\{\hat{\boldsymbol{a}}_i\}$ of mutually commuting operators there exists an operator $\hat{\boldsymbol{b}}$ and a set of functions $\{f_i\}$ such that $\hat{\boldsymbol{a}}_i = f_i(\hat{\boldsymbol{b}})$  (Halmos, 1958). Note, however, that from this mathematical fact it does \textit{not} follow that there also is a \textit{physical measurement} $b$ represented by the operator $\hat{\textbf{b}}$. The existence of such a measurement is a physical question which does not automatically follow from the existence of the operator $\hat{\textbf{b}}$. 

But now suppose that in a KS argument of type III we replace every subset of non-comeasurable measurements $\{a_i\}$ realizing $\{\hat{\boldsymbol{a}}_i\}$ by one single measurement $b$ such that the functions $\{f_i(b)\}$ also realize $\{\hat{\boldsymbol{a}}_i\}$. Will it turn the argument of type III into an argument of type I? 

No, it will not. Replacing non-comeasurable measurements by functions of one single measurement renders the realization hyperedge-based. But then we face the following problem: To test noncontextuality, we need to provide a unique realization of the KS graph and guarantee that all subsets of mutually commuting operators represent simultaneous measurements. However, as Lemma in the Introduction shows, such a realization cannot be hyperedge-based. So we need to give up the uniqueness of the realization, that is, we need to associate at least one operator with more than one measurement. These measurements will be physically different but will be represented by the same operator. Operationally this means that they have the same distribution of outcomes in every quantum state. To get the no-go result, however, one needs to assume more: namely that they have the same distribution of outcomes in every \textit{ontic} state, or in other words, they have the same set of response functions. This assumption, however, is an extra assumption, different from noncontextuality. By abandoned it the KS argument can be blocked.

To sum up, KS arguments of type III do not prove quantum contextuality since FUNC cannot be physically justified for at least one set mutually commuting operators in the argument. Replacing non-comeasurable measurements by functions of one single measurement does not solve the problem either since either we stick to unique realization but then some hyperedges will not represent simultaneous measurements; or we switch to non-unique realization but then we need to use an extra assumption in the argument. To this assumption we turn in the next section.

\section{Spekkens' condition}\label{Sec:Spek}

Rob Spekkens (2005) introduced a constraint on ontological models and called it measurement noncontextuality.\footnote{See also (Liang et al., 2011), (Leifer, 2014) and (Krishna et al., 2017).} He took it to be a generalization of the quantum mechanical noncontextuality for operational theories. I share Spekkens' view that his requirement plays an important role in the KS arguments but, as explained in the Introduction, I contest that it expresses noncontextuality.\footnote{For a criticism of Spekkens operational definition of measurement noncontextuality---based on a criticism of operationalism---see (Hermens 2011).} Hence, I will refer to Spekkens' noncontextuality simply as Spekkens' condition:

\noi
\textit{If the probability of an outcome of a measurement is the same as the probability of an outcome of another measurement in every preparation, then the probability of the outcomes for the two measurements should also be the same in all ontic states.}

\noi
Formally, if for some $x,y \in M$, $k \in K_x$, and $l \in L_y$
\begin{eqnarray}\label{Spekkensnoncontext}
p(X^k|x \w r) = p(Y^l|y \w r) \quad \quad \quad  \mbox{for all}  \, \, r \in S 
\end{eqnarray}
then 
\begin{eqnarray}\label{Spekkensnoncontext2}
p(X^k|x \w \l) = p(Y^l|y \w \l) \quad \quad \quad  \mbox{for all}  \, \, \l \in \Lambda
\end{eqnarray}

Now, Spekkens' condition gives rise to a line of \textit{counterfactual} reasoning. If we measure $x$ in a certain run of the experiment and obtain the outcome $X^k$, then, if the ontological model is value-definite with respect to $x$ and $y$, we can conclude based upon Spekkens' condition that \textit{had} we measured $y$, we \textit{would have} obtained $Y^l$. But note that Spekkens' condition is not an assumption about possible worlds but a restriction on the ontological models for an operational theory. 

Spekkens' condition, similarly to noncontextuality (\ref{mynoncontext}), is also a kind of inference to the best explanation: if (\ref{Spekkensnoncontext2}) and also no-conspiracy (\ref{nocons}) and $\l$-sufficiency (\ref{lsuff}) hold for an ontological model, then we obtain a neat explanation why (\ref{Spekkensnoncontext}) holds. The \textit{explanandum} in the case of noncontextuality is no-disturbance, in the case of Spekkens' condition it is the statistical match between outcomes of different measurements.

Note that Spekkens' condition (\ref{Spekkensnoncontext})-(\ref{Spekkensnoncontext2}) is logically independent from contextuality (\ref{mynoncontext}). Spekkens' condition does not rely on simultaneous measurability, while contextuality does. If there are no simultaneous measurements in an operational theory, then each ontological model will be noncontextual since (\ref{mynoncontext}) is fulfilled vacuously. Still, the model can violate Spekkens' condition (\ref{Spekkensnoncontext})-(\ref{Spekkensnoncontext2}) if there are measurements yielding certain outcomes with the same probability in every state and differing in their response functions. Conversely, if premise (\ref{Spekkensnoncontext}) is not satisfied in an operational theory, then Spekkens' condition is fulfilled vacuously. But if the theory is disturbing, the ontological model can still be contextual. In a non-disturbing operational theory, however, (\ref{Spekkensnoncontext}) holds for all $x$ and $y$ such that $x \geqslant y$. Consequently, if Spekkens' condition holds, noncontextuality will also hold. In short, in a non-disturbing operational theory (like QM) Spekkens' condition implies noncontextuality.

It is instructive to see what an ontological model which violates Spekkens' condition look like. If (\ref{Spekkensnoncontext}) holds in an operational theory   but (\ref{Spekkensnoncontext2}) does not, then the distributions of ontic states representing the preparations cannot be arbitrary. Thus the violation of Spekkens' condition puts a constraint on the possible distributions of ontic states: one cannot pick arbitrarily from ontic states when preparing the system. Preparations must be composed from the underlying ontic states according to a certain pattern which is sensitive to how the ontic states respond to certain measurements. But note that it is not an a priori truth that \textit{any} probability distribution of ontic states represents a physically possible preparation. There may well be many physical reasons which restrict the possible preparations of a system and Spekkens' condition is only one among those.

As we saw in the previous section, Spekkens' condition plays a crucial role in non-unique KS arguments. In these arguments certain operators of the KS graph will be realized by two different measurements. The two different measurements, however---being represented by the same operator---will have the same outcome statistics. But this is exactly the antecedent (\ref{Spekkensnoncontext}) of Spekkens' condition. The role of Spekkens' condition is to ensure the consequent (\ref{Spekkensnoncontext2}), that is, to ensure that the response functions of the two different measurements are perfectly correlated. By this assumption the no-go result can be derived. Thus, non-unique KS arguments heavily rely on Spekkens' condition.\footnote{There are exceptions, however. In certain KS arguments the constraint (\ref{Spekkensnoncontext2}) is not obtained \textit{via} Spekkens' condition but through some other (often counterfactual) reasonings. In Lapkiewicz et al. (2011), for example, an experiment is devised to prove the violation of the Klyachko-Can-Binicioğlu-Shumovsky inequality (2008). To get the conclusion (``to close the pentagram''), however, the authors needed to assume that the response of system on two \textit{not} simultaneous measurements ($A_1$ and $A'_1$ in the paper) are perfectly correlated. This is just a constraint of type (\ref{Spekkensnoncontext2}).}

\section{A simple toy model}\label{Sec:Toy}

Before concluding, it is worth reflecting once more on the difference between noncontextuality and Spekkens' condition  (the first and second interpretations of noncontextuality, as we called them in the Introduction) and illustrating this difference on a simple toy model. Suppose we fill a box with balls and perform two sorts of basic measurements: we pull a ball from the box and check its color or its size. The possible outcomes for the color measurement are black and white; for the size measurement the outcomes are big and small. Repeating the measurement many times we get long-run relative frequencies for the various measurement outcomes. The two measurements are comeasurable, hence also the probability distribution over the joint outcomes can be determined. Suppose furthermore that our operational theory is (i) non-disturbing and (ii) it satisfies the antecedent of the Spekkens' condition: for every preparation, that is, for every filling up the box with balls, the probability of pulling a black ball upon color measurement is the same as the probability of pulling a big ball upon size measurement. 

We would like to construct an ontological model for our operational theory. The model is noncontextual if, given an ontic state, the probability of all four measurement outcomes is independent of whether we produce it by a basic or a joint measurement. The model satisfies Spekkens' condition if, given an ontic state,  the probability of the outcome black/white upon color measurement is the same as the probability of the outcome big/small upon size measurement. 

An ontological model which is both noncontextual and also satisfies Spekkens' condition is the following: there are just two types of balls in the box: one type is black and big, the other type is white and small. Upon measuring the color of the first type of ball we get invariable the outcome black independently of whether we co-measure the size or not (and similarly for the other outcomes). This model neatly explains the above two probabilistic facts, (i) and (ii), of the operational theory.

But there are ontological models in which one of the two requirements is violated. An example of a model satisfying noncontextuality but not Spekkens' condition is the following:  there are now four types of balls in the box: black and big; black and small; white and big; white and small. However, (for some physical reason) we can prepare the box only in such a way that there is exactly as many black and small balls in the box as there are white and big balls. Consequently, although Spekkens' condition is violated, we get as often black balls upon color measurement as big balls upon size measurement.

For an ontological model violating noncontextuality but not Spekkens' condition we need to change our non-disturbing
operational theory into a disturbing one.\footnote{Since, as we saw in the previous section, in non-disturbing
operational theories  Spekkens' condition implies noncontextuality.} Thus, suppose that there are again two types of balls in the box: black and big; white and small. Performing a basic measurement (color, size) these ontic state invariably provide the corresponding outcome. However, for joint measurements  (color \textit{and} size) the outcomes flip: for the ontic state black and big, for example, the outcome for the joint measurement will be white and small. The model is contextual but satisfies Spekkens' condition: the probability of getting a black ball upon color measurement is the same as the probability of getting a big ball upon size measurement in each preparation---both equal to the relative frequency of black and big balls in that preparation. 

As the toy models attest, noncontextuality and Spekkens' condition are different and logically independent assumptions.

\section{Conclusions}\label{Sec:Conc}

In the paper I have argued that a KS argument can rule out a noncontextual value-definite ontological model for QM in a state-independent way only if the KS graph on which the argument is based is (i) given a unique realization such that (ii) mutually commuting operators represent simultaneous measurements. If one abandons (i), then---since some operators will be realized by multiple measurements---one needs to assume Spekkens' condition. By giving up Spekkens' condition, however, the no-go result can be blocked. If one abandons (ii), the constraint FUNC on the value assignments cannot be physically justified. All in all, if noncontextuality is interpreted as the robustness of a system's response to a measurement against other simultaneous measurements, then KS arguments cannot provide an algebraic for proof quantum contextuality.

It is important to note that the main thrust of this negative claim was \textit{not} to challenge the view that QM does \textit{not} admit a noncontextual value-definite ontological model. It does not. State-dependent arguments (like the GHZ argument) provide a perfect proof to this effect. The aim of the paper was to challenge the view that KS arguments can prove this fact \textit{in a purely algebraic way} based exclusively on measurements and not states (and in this sense the KS arguments would be stronger than the state-dependent Bell-type arguments).

But how do we know whether commuting operators represent simultaneous measurements or not? Well, the formalism of QM does not give us a definite answer. One cannot avoid going back and see what kind of measurements the operators are representing. A special way to ensure comeasurability (in a somewhat extended meaning) is to perform the measurements on two or more subsystems of a physical system. These subsystems are typically spacelike separated parts of a bigger system. In the case of spacelike separated measurements noncontextuality (\ref{mynoncontext}) amounts to a locality requirement, called parameter independence: measurements performed on a subsystem cannot influence the response functions of another measurement on a spacelike separated other subsystem.

Noncontextuality as parameter independence plays a crucial role in the Bell-type arguments. In these arguments simultaneous measurability is guaranteed by spacelike separation. KS arguments, however, are not designed specifically against locality but against noncontextuality in general. Therefore, it would be interesting to see whether there exist such KS arguments in which simultaneous measurability is \textit{not} guaranteed by spacelike separation. Obviously, the most baffling form of contextuality is nonlocality. But it would be instructive to see whether there are other ``softer'' versions of contextuality with no appeal to locality. To uncover such contextuality, one should find a family of simultaneous measurements which are performed on the same system (and not on spacelike separated subsystems) and formulate a KS argument based on these measurements. The comeasurability of these measurements should then be justified by explicitly identifying experimental procedures which can be performed on the same system at the same time, like measuring length and width of a table. Such comeasurability would then not appeal to locality but would be justified by the detailed physical description of the measurement processes. Can we come up with a KS argument where comeasurability is grounded in such a way? Does there exist a ``genuine'' KS argument with no appeal to locality? I don't know the answer.

A similarly open question concerns the lack of KS arguments of type I, where all sets of commuting operators represent simultaneous measurements (whether realized by spacelike separation or not). Why are there no arguments providing a state-independent proof for quantum contextuality? Is there a theoretical reason for their non-existence; or are they simply not found because they are not looked hard enough (partly due to the negligence of the difference between commutativity and comeasurability)? Again, I have no answer.

\vspace{0.2in}

\noindent
{\bf Acknowledgements.} This work has been supported by the Hungarian Scientific Research Fund, OTKA K-115593 and a Senior Research Scholarship of the Institute of Advanced Studies Koszeg. I wish to thank the members of the Budapest Research Group on the Philosophical Foundations of Science, especially Márton Gömöri and Balázs Gyenis for valuable discussions and Karim Thebault for reading the final version of the paper.

\section*{References}
\footnotesize

\begin{list} 
{ }{\setlength{\itemindent}{-15pt}
\setlength{\leftmargin}{15pt}}

\item S. Abramsky and A. Brandenburger, ``The sheaf-theoretic structure of non-locality and contextuality,'' \textit{New Journal of Physics}, \textbf{13}, 113036 (2011).

\item A. Acín, T. Fritz, A. Leverrier, and A. B. Sainz, ``A combinatorial approach to nonlocality and contextuality,'' \textit{Comm. Math. Phys.}, \textbf{334(2)}, 533–628 (2015)

\item J. Barrett and A. Kent,  ``Non-contextuality, finite precision measurement and the Kochen-Specker theorem,'' \textit{Stud. Hist. Phil. Mod. Phys.}, \textbf{35}, 151–76 (2004).

\item J. S. Bell, ``On the problem of hidden variables in QM", \textit{Reviews of Modern Physics}, \textbf{38}, 447-452 (1966) reprinted in J. S. Bell, \textit{Speakable and Unspeakable in Quantum Mechanics}, (Cambridge: Cambridge University Press, 2004).

%\item P. W. Bridgman, \textit{The Logic of Modern Physics}, (New York: The Macmillan Company, 1958).

\item A. Cabello, ``A proof with 18 vectors of the Bell–Kochen–Specker theorem", in: M. Ferrero and A. van der Merwe (eds.), \textit{New Developments on Fundamental Problems in Quantum Physics}, (Kluwer Academic, Dordrecht, 1997), 59–62.

\item A. Cabello, S. Severini, and A. Winter, ``Graph-Theoretic approach to quantum correlations,'' \textit{Phys. Rev. Lett}., \textbf{112(4)}, 040401 (2014)

\item R. Clifton and A. Kent, ``Simulating quantum mechanics by non-contextual hidden variables,'' \textit{Proc. Roy. Soc. A}, \textbf{456}, 2101–14 (2000).

%\item L. E. Szabó, ``Mathematics in a physical world,'' in M. Stannet et al. (eds.),  \textit{Pre-proceedings of the 3rd Intentional Hypercomputation Workshop} (HyperNet 11), TUCS Lecture Notes 14, Turku (2011).

\item D. Greenberger, M. Horne, and A. Zeilinger, ``Going beyond Bell’s theorem,'' in M. Kafatos (ed.), \textit{Bell’s Theorem, Quantum Theory, and Conceptions of the Universe} (Kluwer Academic, Dordrecht, 1989), 69–72.

\item C. Held, ``The Kochen-Specker Theorem,'' \textit{Stanford Encyclopedia of Philosophy}, 

URL = https://plato.stanford.edu/entries/kochen-specker/ (2018).

\item P. R. Halmos, \textit{Finite-Dimensional Vector Spaces}, (Dordrecht: Springer, 1958).

\item R. Hermens, ``The problem of contextuality and the impossibility of experimental metaphysics thereof,'' \textit{Stud. Hist. Phil. Mod. Phys.}, \textbf{42 (4)}, 214-225 (2011).

\item A. Klyachko, M. A. Can, S. Binicioğlu, and A. S. Shumovsky, ``A simple test for hidden variables in spin-1 system,'' \textit{Phys. Rev. Lett}. \textbf{101}, 020403 (2008).

\item S. Kochen and E.P. Specker, ``The problem of hidden variables in quantum mechanics", \textit{Journal of Mathematics and Mechanics}, \textbf{17}, 59–87 (1967).

\item A. Krishna, R. W. Spekkens, and E. Wolfe, ``Deriving robust noncontextuality inequalities from algebraic proofs of the Kochen-Specker theorem: the Peres-Mermin square,'' \textit{New J. Phys.}, \textbf{19}, (2017).

\item R. Lapkiewicz, P. Li, C. Schaeff, N. K. Langford, S. Ramelow, M. Wieśniak, and Anton Zeilinger, ``Experimental non-classicality of an indivisible quantum system,'' \textit{Nature} \textbf{474}, 490-493 (2011).

\item J. Larsson, ``A Kochen-Specker inequality,'' \textit{Europhys. Lett.}, \textbf{58}, 799–805 (2002).

\item M. Leifer, ``Is the quantum state real? An extended review of $\psi$-ontology theorems,'' \textit{Quanta.} \textbf{3(1)} 67-155 (2014). 

\item Y. Liang, R. W.Spekkens, H. M. Wisemand, ``Specker’s parable of the overprotective seer: A road to contextuality, nonlocality and complementarity,'' \textit{Phys. Rep.}, \textbf{506 (1-2)}, 1-39 (2011).

\item O. J. E. Maroney and C. G. Timpson, ``Quantum- vs. Macro-Realism: What does the Leggett-Garg Inequality actually test?,''  URL = https://arxiv.org/abs/1412.6139 (2014).

\item M. D. Mazurek, M. F. Pusey, R. Kunjwal, K. J. Resch, and R. W. Spekkens, ``An experimental test of noncontextuality without unwarranted idealizations,'' \textit{Nat Commun.}, \textbf{7}, ncomms11780 (2016).

\item D. Mermin, ``Ontological states and the two theorems of John Bell,'' \textit{Rev. Mod. Phys.}, \textbf{65 (3)},   803-815 (1993).

\item D. A. Meyer, ``Finite Precision Measurement Nullifies the Kochen-Specker Theorem,'' \textit{Phys. Rev. Lett.}, \textbf{83}, 3751–54 (1999).

\item J. L. Park, and H. Margenau, ``Simultaneous Measurability in Quantum Theory,'' \textit{Int. J. Theor. Phys.}, \textbf{1}, 211 (1968).

\item A. Peres, ``Incompatible Results of Quantum Measurements,'' \textit{Phys. Lett. A}, \textbf{151}, 107-108 (1990).

\item M. Reck, A. Zeilinger, H. J. Bernstein, and P. Bertani, ``Experimental realization of any discrete unitary operator,'' \textit{Phys. Rev. Lett}, \textbf{73}, 58 (1990).

\item M. Redhead, \textit{Incompleteness, Nonlocality, and Realism}, (Oxford: Oxford University Press, 1989).

\item C. Simon, C. Brukner, and A. Zeilinger, ``Hidden variable theorems for real experiments,'' \textit{Phys. Rev. Lett.}, \textbf{86}, 4427–4430 (2001).

\item A. Shimony, ``Events and processes in the quantum world,'' \textit{Quantum concepts in space and time}, 182–203,
(1986).

\item R. W. Spekkens, ``Contextuality for preparations, transformations, and unsharp measurements,'' \textit{Phys. Rev. A}, \textbf{71},   052108 (2005).

\item B. C. Van Fraassen, ``Hidden variables and the modal interpretation of quantum theory,'' \textit{Synthese}, \textbf{42}, 155-65 (1979).

\end{list}

\end{document}